\documentclass[useAMS]{mn2e}
\usepackage{graphicx}
\usepackage{times}

\title[Sunspot positions and sizes from Schwabe's observations]
{Sunspot positions and sizes for 1825--1867 from the observations by Samuel Heinrich  Schwabe}
\author[R. Arlt et al.]{R. Arlt$^1$\thanks{E-mail: rarlt@aip.de}, 
R. Leussu$^2$, N. Giese$^{1,3}$, K. Mursula$^2$ and I.G. Usoskin$^{2,4}$
\\
$^1$ Leibniz-Institut f\"ur Astrophysik Potsdam (AIP), An der Sternwarte 16, 
D-14482 Potsdam, Germany
\\
$^2$ Department of Physics, Centre of Excellence in Research, 
University of Oulu, P.O. Box 3000, 90014 Oulu, Finland
\\
$^3$ Kapteyn Astronomical Institute, University of Groningen, Postbus 800, 
9700 AV Groningen, The Netherlands
\\
$^4$ Sodankyl\"a{} Geophysical Observatory (Oulu unit), University of Oulu, 
Finland}
\begin{document}

\date{Accepted \dots. Received \dots; in original form \dots}

\pagerange{\pageref{firstpage}--\pageref{lastpage}} \pubyear{2010}

\maketitle

\label{firstpage}

\begin{abstract}
Samuel Heinrich Schwabe made 8486 drawings of the solar disk with sunspots
in the period from November 5, 1825 to December 29, 1867. We have 
measured sunspot sizes and heliographic positions on digitized images
of these drawings. A total of about 135,000 measurements of individual sunspots 
are available in a data base. Positions are accurate to about 5\% of the solar 
radius or to about three degrees in heliographic coordinates in the solar disk center. 
Sizes were given in 12~classes as estimated visually with circular cursor shapes 
on the screen.  Most of the drawings show a coordinate grid aligned with the
celestial coordinate system. A subset of 1168~drawings have no indication 
of their orientation. We have used a Bayesian estimator to infer the orientations
of the drawings as well as the average heliographic spot positions from a chain of 
drawings of several days, using the rotation profile of the present Sun. 
The data base also includes all information available from Schwabe on spotless days.
\end{abstract}


\section{Introduction}
It is desirable to compile a time series of individual sunspot 
positions going back to the time when telescopes were first used 
to observe them. Such a time series will contain an enormous amount 
of features of great importance for the solar dynamo and the theory 
of magnetic flux emergence at 
the solar surface. A list of existing time series was compiled by Lefevre 
\& Clette (2012). Data of individual  spots have not been available 
for the period  before the Kodaikanal data starting in 1906 until the 
analyses of the Staudacher drawings by Arlt (2009) covering  1749--1799 
and the Zucconi drawings by  Cristo et al. (2009) covering 1754--1760.

The first paper (Arlt 2011, Paper~I) focused on the inventory and description of the 
digitization of the historical sunspot drawings by Samuel Heinrich 
Schwabe made in the period of 1825--1867.
The majority of drawings were made with a high-quality Fraunhofer 
refractor of 3.5~feet focal length.

The full set of 8486~full-disk drawings has now been fully 
measured. The method of measurements will be described in
Section~2 while the resulting spot distribution and the possible
sources of errors will be discussed in Section~3. The
analysis aims at the full exploitation of the drawings
by providing positional information of each individual 
sunspot together with its size. Unfortunately, the Greenwich
data set and its continuation by the USAF/NOAA only provides
the average group positions and the total areas of the groups.
Information like the size distribution of sunspots and the
tilt angles and polarity separations of bipolar regions are
only preserved if the individual spots are stored in the
data set, however.

The Schwabe data are also superior to the ones by Carrington
(1853--1861; cf.\ Lepshokov et al. 2012 for a recent analysis) 
and Sp\"orer (1861--1894; recent analysis by Diercke et al. 2012),
which only report about sunspot groups at a certain instance
when they were near the central meridian. The Schwabe data contain
the full evolution of sunspot groups crossing the visible
solar disk.

\section{Methods of measurements}
\subsection{Heliographic coordinate system}
For all images possessing a horizontal reference line, we assumed
that the line is parallel to the celestial equator (cf.\ Paper~I).
The position angle and tip angle of the heliographic coordinate
system is obtained from the JPL Horizons ephemeris generator\footnote{http://ssd.jpl.nasa.gov/horizons.cgi}. We used the geographical coordinates of the observing location in the town of Dessau, Germany, and generated a list of these quantities in six-hour intervals for the entire period of 1825 to 1867. The quantities for times in between two output lines were  interpolated linearly.
The documentation of the Horizons ephemeris service states  that the position 
angle is the ``target's North Pole position angle (CCW with 
respect to direction of true-of-date celestial North)''. It is reasonable to assume that Schwabe used the local sky rotation to adjust his telescope to the north. He must thus have been arrived nearly at a ``true-of-date'' celestial north. The actual orientation of the solar-disk drawing comes from the cross-hairs used in the eyepiece. Schwabe did not report on how he adjusted the eyepiece (rotation may have easily been possible). Throughout the vast majority of observations the alignment is amazingly consistent but, as we will see later, there are a few short periods when the eyepiece was apparently misaligned.

For all observing days  with drawings, the actual solar disk was extracted from the digitized image by four
mouse clicks on the left, right, lower and upper limbs of the
circle, where the middle of the pencil stroke width was chosen. This way,
also slight ellipticities are allowed thereby, although only 
in the vertical or horizontal directions and not at an arbitrary
angle. This turned out to be a reasonable choice, since ellipticities
mainly come from the fact that the paper may not have been entirely 
flat when photographed, producing a prolateness of the circles.

If a horizontal line is available in the image, two clicks near
the left end and near the right end of the line define the 
position angle of the celestial equator in the image. Again, the middle 
of the pencil stroke width was chosen visually. The position 
angle of the solar equator is then added to this orientation, and
the actual heliographic grid is superimposed to the image.

In some cases, the main vertical line is not perpendicular to the
horizontal one. We are applying a special transformation to the
cartesian coordinates of the measurements, as described in Subsect.~\ref{skew}
below. The various tools for the measurements were written in the 
Interactive Data Language (IDL).
 
\subsection{Method for unoriented drawings\label{bayesian}}
There is a set of 1168~drawings which do not show a coordinate 
system, mostly in the period of mid-1826 to 1830. Since there are
often sequences of days for which the drawings have a number of
sunspots in common, we can use the rotation of the Sun to find
the probable position angles of the heliographic coordinate systems.
We assume the sidereal rotation profile obtained from {average sunspot 
group positions} by Balthasar et al. (1986) and use the numerical values
\begin{equation}
  \Omega(b) = 14.551 \degr/{\rm d} - 2.87 \degr/{\rm d} \sin^2 b
  \label{rotation_profile}
\end{equation}
for the angular velocity $\Omega$, where $b$ is the heliographic
latitude. We actually need the synodic rotation rate for our purposes
which is obtained from solving Kepler's equation for the eccentric anomaly
of the Earth at each instance it is needed, using an eccentricity
of $e=0.01687$, and a rotation period of $P_{\rm rot} = 365.24219879~{\rm d}
-6.14\cdot 10^{-1} (JD - 2415019)/36525$ (Newcomb 1898).
Note that the use of the solar rotation profile implies that we cannot use the
resulting sunspot positions directly for the determination of the 
differential rotation of the Sun later on, since they are not 
independent of the rotation profile.

A Bayesian parameter estimation is employed to obtain the position
angles and average sunspot positions. We start with looking at $n_{\rm d}$ drawings and associating
$n_{\rm s}$ sunspots  with each other, which are visible in all these
drawings. Given the two coordinates of each spot, these combinations deliver  
$N=2n_{\rm s}\,n_{\rm d}$ measurements. The unknowns are the heliographic
coordinates of the spots, $l_i$ and $b_i$, where $i=1,\dots,n_{\rm s}$ 
counts the spots, and the position angles $p_j$ of the drawings where 
$j=1,\dots,n_{\rm d}$ counts the days.
We are thus faced with $M = 2n_{\rm s}+n_{\rm d}$ free 
parameters. 
%Table~\ref{unknowns}
%illustrates the number of unknowns versus the number of measurements for
%various typical combinations possible from Schwabe's observations.
For three days with three common spots, we have $N=18$~measurements and $M=9$~unknowns,
for example, while two days with two spots deliver only $N=8$
and $M=6$. Note that there may be two or three days between two adjacent
drawings in a sequence.

Formally, there is another parameter which we either have to 
determine beforehand or keep as a free parameter. It is the 
measurement error of Schwabe's plots. It is reasonable to assume that 
these errors roughly form a Gaussian distribution. Deviations from
Gaussian distributions may only be expected for spots very near the
solar limb, but for the majority of spots, Gaussian will be a good
approximation, and we assume that there is a single standard deviation
$\sigma$ describing the distribution. Allowing $\sigma$ to be
a free parameter was considered, but turned out to be impractical since
the model then obtains excessive freedom to assume that the spots are
in the wrong place and yield very odd combinations of latitudes 
and position angles at high likelihood. The value of
$\sigma$ was thus estimated from a number of chains with high
$n_{\rm d}$ and high $n_{\rm s}$ using the residuals. These 
should be identical to the plotting errors only for infinitely 
large $n_{\rm d}$ and $n_{\rm s}$, an exactly known rotation 
profile and the assumption of zero proper motion of the spots. As 
a compromise we chose chains of five drawings having 2--4~spots in 
common and kept $\sigma$ as a free parameter. From this set of 15~sample 
chains (i.e. 75~drawings) in 1827 and 1828, we  obtained an a average 
$\sigma=0.05$ of the solar disc radius. We used this value of $\sigma$ in all actual determinations 
of position angles of drawings where no coordinate  system  was given by 
Schwabe. Note also the additional remarks about the
accuracy in Section~\ref{accuracy}.

%\begin{table}
%\caption{Comparison of the number of measurements versus the number 
%of free parameters for given numbers of drawings and spot numbers. }
%\label{unknowns}
%\begin{tabular}{llcc}
%Drawings, $n_{\rm d}$ & Spots, $n_{\rm s}$ & Measurements & Parameters, $M$\\
%\hline\\
%2 & 2 & 8  &  6\\
%2 & 3 & 12 &  8\\
%2 & 4 & 16 & 10\\
%2 & 5 & 20 & 12\\
%2 & 6 & 24 & 14\\
%2 & 7 & 28 & 16\\
%3 & 2 & 12 &  7\\
%3 & 3 & 18 &  9\\
%3 & 4 & 24 & 11\\
%3 & 5 & 30 & 13\\
%3 & 6 & 36 & 15\\
%3 & 7 & 42 & 17\\
%4 & 2 & 16 &  8\\
%4 & 3 & 24 & 10\\
%4 & 4 & 32 & 12\\
%4 & 5 & 40 & 14\\
%4 & 6 & 48 & 16\\
%4 & 7 & 56 & 18\\
%5 & 2 & 20 &  9\\
%5 & 3 & 30 & 11\\
%5 & 4 & 40 & 13\\
%6 & 1 & 12 & 8 \\
%6 & 2 & 24 & 10 \\
%6 & 3 & 36 & 12\\
%6 & 4 & 48 & 14\\
%\hline
%\end{tabular}
%\end{table}

Bayesian inference is based on the distribution of probability density
over the entire parameter space. (We will often use the term `probability
distribution', but actually the probability density is meant.) Every combination of parameters,
given the model differential rotation, is tested on its likelihood
to have created the data. Since this is too expensive computationally,
we are employing Monte-Carlo Markov chains which explore the parameter
space very efficiently, without wasting computing time in regions
of very low probability density but without being limited to local maxima either. The  parameter space for the determination of orientations from one chain is binned into
$2048^M$ bins for which the number of passages of the Markov chains
is counted. After normalisation, these counts give the probability density
distribution. The posterior distribution for a given individual parameter
is obtained by marginalisation over all the other parameters.

One often has several options of combining consecutive drawings
into a chain that is analyzed by the Bayesian estimator. It is of 
course not a matter of the residuals to tell which combination
is best, since the residuals always improve when the number of
free parameters approaches the number of measurements. We will denote
the combinations by $n_d\backslash n_s$ in the following.

The suitability of combinations of drawings for the determination of the orientations is not easily quantified. The Bayesian Information Criterion (BIC, also called Schwartz Criterion) is one guess for the trade-off between keeping the residuals as well as the number of free parameters low. It does not, however, take into account the distribution of the spots over the solar disk which may vary from very suitable to almost degenerate. We computed a number of about twenty test cases to obtain an idea of good and bad distributions. Based on the BIC and on this experience, we start from the combination $3\backslash3$ 
as the desired one and used a ranking for other combinations to be 
chosen when $3\backslash3$ is not possible. The ranking with descending ``priority'' is the following: $3\backslash3, \, 2\backslash6, 2\backslash5,\, 3\backslash4,\, 4\backslash2,\, 2\backslash4,\, 2\backslash3$ and $3\backslash2$ (turned out to be equal  in suitability), $2\backslash2, \, 6\backslash1$. Rare occasion with many common spots are not listed here, but the first three combinations indicate already, that it is better to use three drawings with few  spots than a pair of drawings with many spots. Since there is considerably more rotational displacement with three drawings, the third drawing fixes the spot positions very well and always produces highly plausible results.

A subjective quality flag is given to the spots of a given day.
All drawings with a pencil grid obtain a quality flag of $Q=1$.
Positions derived from the rotational matching of two or more
consecutive drawings, may obtain $Q=1$, $2$, or $3$. Our rules 
to assign the different quality flags are as follows: if the 
probability distribution of one of the free parameters has a full 
confidence interval of more than $40^\circ$, but  the distributions
are not skewed, we assign a quality of $Q=2$. If the distributions are
additionally slightly skewed so that the average parameter is different 
from the mode value by up to $20^\circ$, we use $Q=3$. A subjective estimate of the quality is given with values from 1 (highest quality) to 3 (lowest quality). Drawings 
delivering very skewed or double-peaked distributions are discarded and the quality estimate is set to 4 (see Section~2.7). 
We did not derive any sunspot positions  for those days. We  store the sunspots of discarded 
drawings and fill their positions with NaN. Yet their sizes are 
available and useful and are stored in the spot file along with 
group designations. 

It is of great advantage to know the full distribution of the probability density
as compared to minimization procedures for, e.g., $\chi^2$. Such
searches are not aware of additional minima and may even miss the
global minimum entirely.

\begin{figure}
\centering
\includegraphics[width=0.48\textwidth]{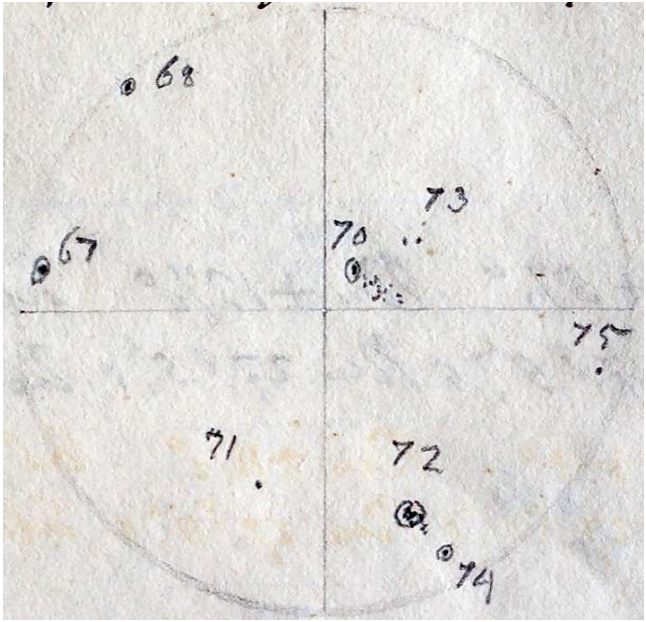}
\caption{Example drawing of 1836 April 11 with penumbrae. Most of Schwabe's
drawings are made in this style.\label{penumbrae}}
\end{figure}

\subsection{On-screen measurements}
We used a set of  circular mouse cursor shapes with different diameters to estimate the sizes and
positions of the sunspots. For all spots showing a penumbra,
only the umbral size was measured. This is because open circles were often drawn 
 by Schwabe  to indicate the presence of penumbrae. While the umbrae pencil dates are clearly drawn with the intention to distinguish different sizes, the penumbrae show less carefulness since they are all of very similar size. Additionally, Schwabe's penumbrae show little foreshortening near the solar limb (see group~68 in Fig.~\ref{penumbrae}). We leave it to future scrutinization which may or may not show the scientific usefulness of the penumbral sizes drawn by Schwabe.

A total of 12~size steps of a circular cursor were used 
running from an area of 5~square-pixels to 364~square-pixels (Table\ref{sizes}) including the borders. We always used the 
largest possible circular cursor  for which the boundary of the circle was
contained {\em within\/} the umbral area, if the umbra was circular.
Noncircular spots can only be approximately matched with these  cursor masks, of course. Note that the pencil dots have a certain minimum
size which did not require the use of 1~square-pixel areas. The total area
of the solar disk is 708822~pixels. A single square-pixel corresponds to
1.4~millionths of the disk. The smallest areas measured here are
7~millionths of the solar disk. An alternative way of estimating the areas
was given by Cristo et al. (2011). In their
work, the umbral areas were derived for Zucconi's observations in 
1754--1760 in a semi-automatic black-pixel-finding algorithm which can
deal with almost arbitrary sunspot shapes. Because of the lower and
varying contrast in Schwabe's pencil drawings,  this 
algorithm would be more difficult to apply in our case, and was not employed.

Before 1831, Schwabe did not distinguish umbra and penumbra in 
his drawings. The first full-disk drawing with distinguished penumbrae
is from 1831  Jan~06. At the same time, Schwabe stopped
drawing magnifications of sunspot groups besides the full-disk drawings on a regular basis and did so only for
spectacular groups or interesting observational facts he wanted to 
emphasize. We will have to choose an appropriate calibration for the
sunspot areas in order to obtain a consistent data set.

\begin{table}
\caption{Cursor sizes and corresponding areas in square-pixels.}
\label{sizes}
\begin{tabular}{rr}
\hline
Size  &  Area \\
\hline
1     &  5 \\
2     &  9 \\
3     & 21 \\
4     & 37 \\
5     & 69 \\
6     & 97 \\
7     & 145\\
8     & 185\\
9     & 206\\
10    & 270\\
11    & 308\\
12    & 364\\
\hline
\end{tabular}
\end{table}

%\begin{table}
%\caption{Cursor sizes and corresponding areas in square-pixels.}
%\label{sizes}
%\begin{tabular}{rr|rr|rr}
%\hline
%Size  &  Area & Size  &  Area &Size  &  Area \\
%\hline
%1     &  5   &  5  & 69 & 9  & 206\\
%2     &  9   &  6  & 97 & 10 & 270\\
%3     & 21   &  7  &145 & 11 & 308\\
%4     & 37   &  8  &185 & 12 & 364\\
%\hline
%\end{tabular}
%\end{table}

We did not contemplate using elliptical cursor shapes for foreshortened
sunspots near the solar limb. The cursor size was chosen visually as to
approximate the roughly elliptical shape of the sunspot by a circle of 
equal area instead, but still referring to the projected sunspot area. The introduction of different ellipticities for
different limb distances would have made the measurements considerably
more time-consuming.

For the sunspot position the appropriate cursor shape was centered on the pencil dot in the
image visually and the position fixed by a mouse click. We decided
to use only the spots visible in the full-disk drawings, delivering a
consistent set of spots always drawn at the same scale. Detailed 
drawings of sunspot groups next to the full-disk drawings were not used despite containing additional fine pores.

All positions were first stored in a momentary reference frame with the 
$0^\circ$-meridian running through the center of the disk (central-meridian
distance, CMD). If the interpretation of the times given by Schwabe should change,
new Carrington longitudes could always be generated from the momentary
reference frames.

\begin{figure}
\centering
\includegraphics[width=0.48\textwidth]{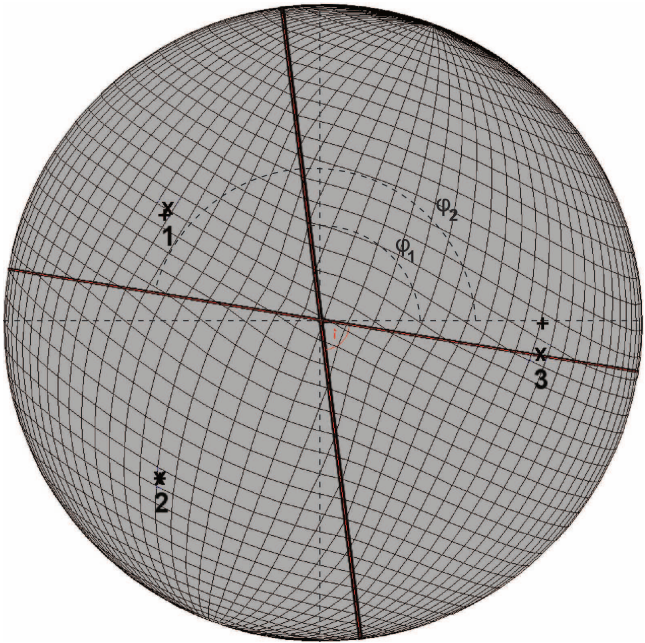}
\caption{Highly exaggerated test case for the correction of skewed
coordinate systems. The x-shape symbols represent spots in the drawing, the +-symbols are the corrected positions.
{The angles $\phi_1$ and $\phi_2$ are used in Eq.~(\ref{skewmap})}.
\label{distorted}}
\end{figure}

\subsection{Skewed coordinate systems\label{skew}}
The main vertical and horizontal lines are not always
perfectly perpendicular. In cases where the difference from 
$90^\circ$ is more than $1.66^\circ$ {(corresponding to 
roughly half the plotting accuracy -- see Sect.~\ref{accuracy})}, 
we applied a transformation to the normalized, cartesian coordinates
before we converted them into heliographic ones.

Since it is the lines on paper that have to be drawn anew
every day, while the actual cross-hairs in the eyepiece
need no re-alignment, we assume that the
eyepiece was correct, whereas the drawing was imperfectly
made.

When copying the visual information on the spot positions
from the eyepiece, the lines were used as references. 
If the lines on paper differ from the view in the eyepiece, the (additional)
plotting error is the larger the closer the spot is to one 
of the reference lines. The spots near any of the lines will be offset by the same amount as the reference line is offset against the real view in the telescope.

Let us consider a polar coordinate system with the intersection 
between the `horizontal' and `vertical' reference lines being the origin. Any 
spot will appear in a sector between such `horizontal' and 
`vertical' lines. Let $\phi_1$ and $\phi_2$ be the two angles 
at which these `horizontal' and `vertical' lines are drawn. We
also convert the measured cartesian $(x,y)$ into polar coordinates
$(r,\phi)$ on the solar disk. {The angles $\phi_1$ and $\phi_2$ 
as well as the cartesian and polar coordinates are defined in the 
usual rectangular coordinate system aligned with the image 
coordinates which is only of auxiliary nature. The situation is 
depicted in Fig.~\ref{distorted} where the deviation from perpendicularity
is exaggerated for clarity. The correct Schwabe system is now 
positioned in such a way that the new lines} have equal angular distances
from the plotted `horizontal' and `vertical' lines, respectively, and
are perpendicular {(not plotted in Fig.~\ref{distorted})}. This 
angular distance is denoted by $\alpha$. We correct the spot position by
\begin{equation}
  \phi' = \phi + \alpha\left(\frac{2\phi}{\phi_2-\phi_1}-1\right)^q
  \label{skewmap}
\end{equation}
where the new location is $(r,\phi')$. $\alpha$ is the deviation of 
the vertical and the horizontal lines from being rectangular, 
$\alpha=(\phi_2 - \phi_1 - 90^\circ)/2$.  The term in {parentheses in Eq.~\ref{skewmap}} gives
numbers between $-1$ and $1$ which are multiplied by the maximum
shift which would be necessary if the spot is exactly on one of 
the wrong axes at $\phi_1$ or $\phi_2$. The exponent $q$ controls the strength of the re-mapping. {A small $q$ causes} the re-mapping to be effective over most of the sector between a horizontal and a vertical line. A large $q$ causes the re-mapping to be confined close to the lines while being practically zero in the ``field'' between the lines. We used $q=1$ throughout the analysis.

\subsection{Typical problems occurring}
All measurements are made manually. This allowed us to interpret
what is meant in the drawing at every instance of the process.
Some features in the images can mimic sunspots and need to be distinguished.
\begin{itemize}
\item Paper defects. They usually  have a slightly brownish
  colour and can be distinguished from pencil-drawn sunspots
  quite easily.
\item Faculae were often marked in the drawings, but of course
  not as bright features but with weak, often curved pencil strokes.
  Visual inspection often tells what are faculae in a given group 
  and what are small spots. Faculae without spots (especially
  near the solar limb) do not have group numbers and can thus be
  omitted from the measurements. In doubtful cases, the verbal descriptions
  can be used as they regularly report on the presence of 
  faculae (``Lichtgew\"olk'').
\item Dots associated with group numbers. Schwabe often added a dot
  after the group number to mark it as an ordinal number. Also the
  number 1 gets a top dot to distinguish it from a simple vertical 
  line. Hence, group number 11 comes along with two additional dots
  in the drawing. They are drawn in ink and appear darker than
  the pencil-drawn sunspots.
\item Pinholes from the pair of compasses. While the pinhole of the
  actual drawing is  obviously not disturbing as it is passed by the vertical and horizontal reference lines, pinholes from drawings on the back of the paper can appear anywhere in the solar disk. They can be distinguished from spots since they exhibit a raised appearance in contrast to the engraved pencil dots of the real spots.
\end{itemize}

\begin{figure}
\centering
\includegraphics[width=0.48\textwidth]{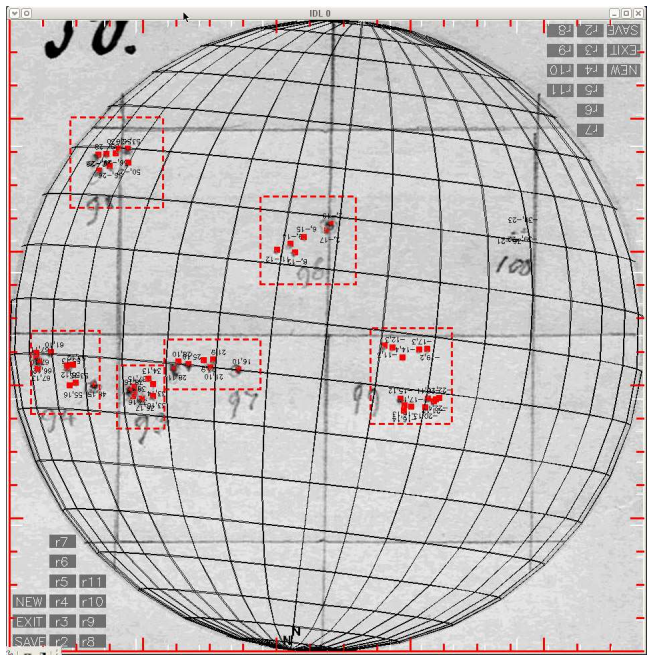}
\caption{Screen-shot of the group numbering tool for the drawing
of 1861 Jun 22. The actual measurement was made with an image rotated
by $180^\circ$, since Schwabe's drawings are all upside-down. This
is why the coordinates are upside-down, while it is more convenient
to use Schwabe's original orientation for reading the group numbers.
The picture is a screen-shot from the process of numbering, whence
the yet unnumbered group 100.
\label{numbering}}
\end{figure}

\subsection{Group numbers}
Schwabe numbered the groups starting with number one each year. A few
groups visible already in the previous year carried their numbers into
the new year. Schwabe tried to identify groups from previous solar
rotations when they became visible again. He mentioned possible 
re-apparitions but always assigned new numbers to any group appearing on
the eastern limb of the Sun.

We store the group designations for each spot measured. They are not
always numbers. In the very beginning, Schwabe used letters. Faculae
-- most prominently visible near the solar limb -- were often referred
to by Greek letters. When Schwabe referred to parts of a group in his
verbal descriptions, he also used Greek letters very often.

Note that the definition of a group is not necessarily identical
to a group definition we would use today. Two bipolar groups at 
the same heliographic longitude but at slightly different latitudes
were most likely classified as  a common group  although we would separate
them as two groups with today's knowledge. Another difficulty
arises from the foreshortening when new groups appear near the limb.
Schwabe assigned a single group number to some sunspots appearing
at the limb, although they turn out to be two or more groups when
the full longitudinal extent becomes evident in the middle of the
solar disk. An example of the numbering is given in Fig.~\ref{numbering}.
While the numbering is typically fine, we also see an example 
(group no.~99) where two groups were combined into one group.
Nevertheless, we kept the original group numbers to preserve
as much of the historical information of the drawings as possible.

%and (ii) to have a quick method to evaluate the plotting accuracy
%of Schwabe. It will be easy to track the mean latitudes of a given
%group as it passes the solar disk and study their variation (cf.\
%Section~\ref{accuracy}. 

\begin{table*}
\caption{Data format of the data base of sunspot observations by
Samuel Heinrich Schwabe for the period of 1825--1867. The fields are separated
by one blank space each which is not included in the format declarations.\label{format}}
\begin{tabular}{lllp{11cm}}
\hline
Field & Column & Format & Explanation \\
\hline
Year  & 1--4   & I4     & Year \\
Month & 6--7   & I2     & Month \\
Day   & 9--10  & I2     & Day referring to the German civil calendar running from midnight to midnight\\
Hour  & 12--13 & I2     & Hour, times are mean local time \\
Minute& 15--16 & I2     & Minute, typically accurate to 15~minutes\\
Timeflag& 18   & I1   & Indicates how accurate the time is. Timeflag\,=\,0 means the time has
                 been inferred by the measurer (in most cases to be 12h~local time);
                 Timeflag\,=\,1 means the time is as given by the observer\\
L0    & 20--24 & F5.1   & Heliographic longitude of apparent disk center seen from Dessau\\
B0    & 26--30 & F5.1   & Heliographic latitude of apparent disk center seen from Dessau\\
CMD   & 32--36 & F5.1   & Central Meridian Difference, difference in longitude from disk center; 
                 contains --.- if line indicates spotless day;
                 contains NaN if position of spot could not be measured.\\
Longitude&38--42&F5.1 & Heliographic longitude in the Carrington rotation frame; 
                 contains --.- if line indicates spotless day; 
                 contains NaN if position of spot could not be measured.\\
Latitude&44--48& F5.1 & Heliographic latitude, southern latitudes are negative; 
                 contains --.- if line indicates spotless day; 
                 contains NaN if position of spot could not be measured.\\
Method  & 50   & C1   & Method of determining the orientation. `C': horizontal pencil line parallel to 
                 celestial equator; `H': book aligned with azimuth-elevation; `Q': rotational 
                 matching with other drawings (spot used for the matching have ${\rm ModelLong}\neq {\rm `-.-'}$, 
                 ${\rm ModelLat}\neq {\rm `-.-'}$, and ${\rm Sigma}\neq {\rm `-.-'}$).\\
Quality & 52   & I1   & Subjective quality, all observations with coordinate system drawn by Schwabe get 
                 Quality\,=\,1, also the ones with skewed systems that were rectified by the method described in Section~\ref{skew}.
                 Positions derived from rotational matching may also obtain Quality\,=\,2 or 3, if the
                 probability distributions fixing the position angle of the drawing were not
                 very sharp, or broad and asymmetric, respectively. Spotless days have Quality\,=\,0; 
                 spots for which no position could be derived, but which have sizes, get Quality\,=\,4.\\
Size    & 54--55   & I2   & Size estimate in 12 classes running from 1 to 12; a spotless day is indicated by 0 \\
SGroup  & 57--64   & C8   & Group designation taken from Schwabe. \\
Measurer& 66--75   & C10  & Last name of person who obtained position\\
ModelLong & 77--81 & F5.1 & Model longitude from rotational matching (only spots used for the matching have this)\\
ModelLat & 83--87  & F5.1 & Model latitude from rotational matching (only spots used for the matching have this)\\
Sigma    & 89--94  & F6.3 & Total residual  of model positions compared with measurements of reference spots in rotational matching (only spots used for the matching have this). Holds for entire day.\\
\hline
\end{tabular}
\end{table*}

\section{Description of the data file}
The data are arranged in a format described in Table~\ref{format}.
There is a single blank space between each of the data fields. The
first five fields contain the time to which the positions refer.
It is fairly certain that the times of observations are mean local 
times, since Schwabe made efforts to determine culmination
times and keep track of deviations of his clocks of the order of
seconds. In some cases, the time to which the full-disk drawing 
refers is ambiguous or missing. When missing, we assumed 12h~local 
time and set the ${\tt Timeflag}=0$ for these cases. When several 
times were given ambiguously, we used the most probable time given -- in most
cases also 12h as the days before and after are typically stating
12h clearly for the times of the drawings.

The columns {\tt L0} and {\tt B0} are the heliographic coordinates of the center 
of the Sun as given by the JPL Horizons ephemeris service (`Observer sub-long \& sub-lat').
The coordinates are for the apparent disk center as seen from the 
observing location in Dessau, but the differences to the topocentric
coordinates are far below the plotting accuracy (parallax of
order $0\fdg002$). L0 and B0 are equal for all spots of a given
day, of course. But we give them for every spot to ensure the
conversion to the Carrington frame of heliographic coordinates 
will be replicable. Note that Horizons obtains a
zero longitude for the disk center on 1853 Nov~9, at $21^{\rm h}36^{\rm m}$~UT.
Carrington defined the zero point of his longitude counting on
1853 Nov~9.

The {\tt CMD} is the central-meridian distance and is a heliographic longitude
measured from the central meridian where values west of it (seen on
the observer's sky) are positive, and values east of it are negative.
The direction of measuring longitudes is therefore the same as 
Carrrington's. Heliographic longitudes in the Carrington frame are then
obtained by adding L0 to CMD. The final Carrington coordinates are
stored in the columns {\tt Longitude} and {\tt Latitude}.

The {\tt Method} field contains a character denoting the method by
which the orientation of the solar disk was obtained. The
most frequent value is `C' which stands for celestial
system. The main horizontal line in the drawing was assumed
to be parallel to the celestial equator. The orientation of
the heliographic system is based on this assumption. The character
`Q' stands for a rotational matching described in Subsection~\ref{bayesian}.
The character `H' denotes observations without lines, for which
we assumed that the orientation of the book is parallel to the
horizon. If the observation was made at noon, this is equal
to being parallel to the celestial equator. The apparent
rotations of the disk drawn led us to the conclusion that 
disks at other times of the day are not oriented in a
celestial, but rather in a horizontal system.

The {\tt Quality} field gives a subjective quality of the positions
on a scale from~1 to~3.
All drawings with a pencil-drawn coordinate system obtained a 
Quality of~1. All drawings for which the Method is `H' obtained
a Quality of~3. Drawings treated by rotational matching obtain
a Quality of 1 for narrow probability distributions, a Quality
of 2 for broad, but symmetric probability distributions, and a
Quality of 3 for skewed probability distributions. Note that
the quality flag only refers to the accuracy of the positions,
not the spot sizes.

The values for {\tt Size} are given from the original measurement. A
conversion of these size classes into, e.g., micro-hemispheres is
difficult and needs to be done at a later stage of comparing the
Schwabe data with other sources. Spots were plotted as simple 
pencil dots of various size until 1831, while the first 
distinction between umbra and penumbra was made on 1831 Jan~06
and continued to be made throughout the rest of the observations.

Foreshortened spots near the solar limb where usually plotted
as elliptical dots. In principle, our size estimates are projected areas; 
we tried to use a circular cursor shape which has an area equal to the 
elongated spot plotted. Given the difficulty in drawing arbitrarily
thin lines with a pencil, however, we have to assume that these projected areas
are overestimated as compared to the spot sizes near the disk
centre.

In case of days without sunspots, there is a single line in the data file
with `-.-' in the sunspot position, while we set ${\rm Size}=0$.
Note that even then, we cannot provide a full
record of Schwabe's observations, since many of the 3699~verbal reports cannot
be represented in this data format. The reports of spotless days are all
incorporated in the data base with lines having ${\rm Size}=0$, while the remaining reports may be utilized in a future step of analysis of Schwabe's observing records. There is usually only information on the appearance of new, or disappearance of existing groups, compared to the previous observation. Group sunspot numbers may easily be determined for these days, but only by assuming Schwabe's definition of a group is correct (or compatible with our today's understanding). It will also be possible to improve the group  sunspot numbers by Hoyt \& Schatten (1998) according to the verbal reports.

The column {\tt SGroup} contains the group designation given by Schwabe.
The {\tt Measurer} column gives the last name of the person who obtained the spot position. The full names can be retrieved from the list of authors and the acknowledgments.

\begin{figure*}
\centering
\includegraphics[width=0.98\textwidth]{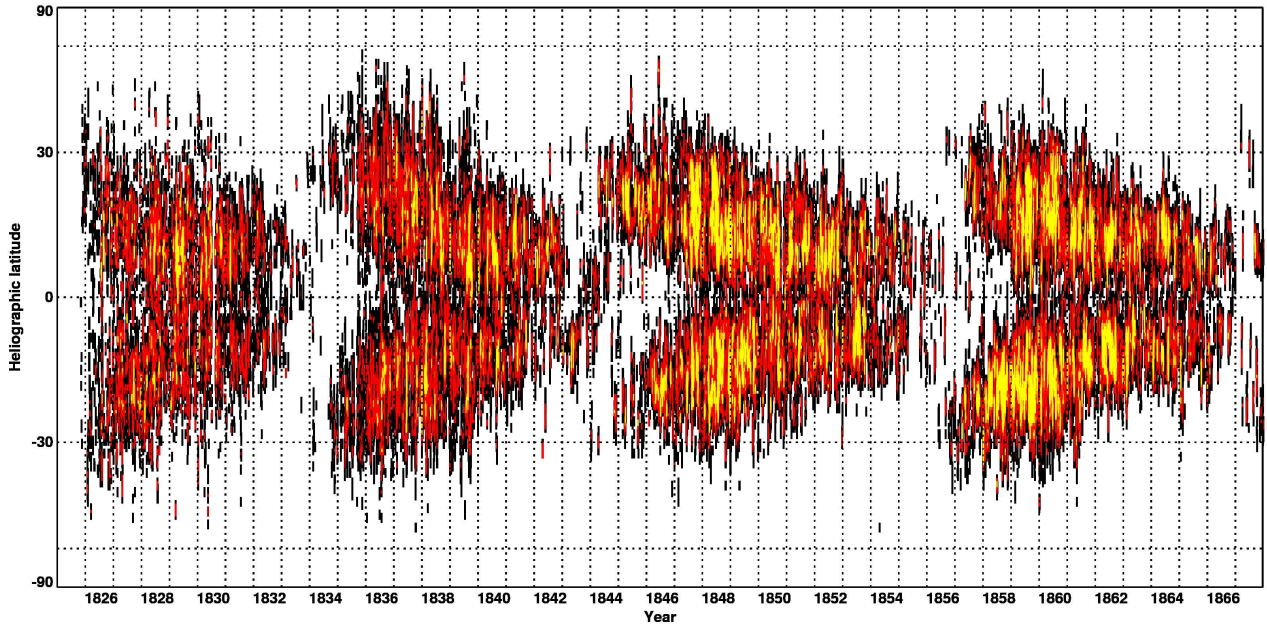}
\caption{Butterfly diagram based on about 135,000~sunspot positions derived
from Schwabe's observations of 1825--1867. A similar plotting style
as used by Hathaway\protect\footnotemark[2]\ is employed here.  \label{butterfly}}
\end{figure*}

Additional information is given for the drawings that were analysed using the rotational matching. The spots used to fix the orientations of the drawings deliver posterior distributions for their positions as a side-product. We computed the averages of these posterior distributions and added the resulting positions to the corresponding lines in the data base as {\tt ModelLong} and {\tt ModelLat}. Since the model assumes stationary spots, the latitudes of the spots are constant for the drawings involved in this particular rotational matching. The longitudes are not exactly constant because they are Carrington longitudes, and the spots drift against the Carrington frame of reference according to the rotation profile (\ref{rotation_profile}) used. The {\tt Sigma} column contains the standard deviation of the spots involved in the matching.

Occasionally, the model position does not refer to exactly the spot it is attached to. This results from spots that had split during the course of the period used for the rotational matching. The model position was then compared with the middle of the two new spots while the actual measurement afterwards, with the inferred orientation, generated two lines in the data base for the two spots.

\begin{table}
\caption{Sample series for an estimate of the plotting accuracy\label{sigmas}}
\begin{tabular}{l@{~}r@{--}l@{~}rrr}
\hline
\multicolumn{4}{l}{Period}  & Days & $\sigma$ \\
\hline
1832 Feb&01&Feb&04 &  4 & $2\fdg13$ \\
1832 Feb&10&Feb&15 &  4 & $3\fdg75$ \\
1832 Feb&15&Feb&25 & 11 & $2\fdg56$ \\
1832 Feb&24&Mar&06 &  6 & $2\fdg54$ \\
1845 Jan&12&Jan&24 &  7 & $4\fdg48$ \\
1845 Feb&13&Feb&24 &  6 & $5\fdg06$ \\
1845 Jul&06&Jul&10 &  4 & $0\fdg91$ \\
1854 Apr&06&Apr&14 &  7 & $2\fdg12$ \\
1855 Mar&06&Mar&18 &  8 & $3\fdg08$ \\
1855 Oct&23&Nov&02 &  8 & $1\fdg46$ \\
1856 Apr&10&Apr&20 &  9 & $1\fdg20$ \\
1864 Jan&12&Jan&23 &  8 & $2\fdg07$ \\
1864 Jul&01&Jul&12 &  7 & $2\fdg95$ \\
1865 Feb&05&Feb&14 &  5 & $3\fdg92$ \\
1865 Jul&09&Jul&19 &  8 & $1\fdg59$ \\
\hline
\multicolumn{5}{l}{Simple average}   & $2\fdg65$ \\
\hline
\end{tabular}
\end{table}

\section{Spot distribution and accuracy of the drawings}\label{accuracy}
As already discussed in Section~\ref{bayesian}, the analyses of 75~drawings 
{\em without\/} reference lines delivered a plotting accuracy of 0.05 in units of the 
solar radius ($2\fdg9$ in the disk center). We might consider this an upper limit,
since the absence of reference lines makes accurate plotting more difficult, but
see below.

The plotting accuracy certainly varied on a day-by-day scale, since 
poor weather may have allowed only little time for a drawing. This 
is supported by occasional comments by Schwabe  that the spot positions 
are only approximate because of clouds.

For the accuracy of the majority of drawings which do show a
coordinate grid, we selected blindly a number of sequences
of days during which simple spots of Waldmeier class~H and~I were
crossing the solar disk. We determine the deviations of the measured
latitudes from the average latitude of such a spot. To avoid problems with the differential
rotation, we only looked at the scatter in the heliographic
latitudes and assume the true latitudes have not changed with time. 
The periods with the resulting standard deviations $\sigma$
of the spots' latitudes are given in Table~\ref{sigmas}. 
The average $\sigma$ ($2\fdg65$)  needs to be converted into
a total angular error, since we have only considered the
latitudes here, whence approximately $\sigma_{\rm tot} \approx
\sqrt{2}\sigma = 3\fdg75$. Interestingly, this value is even a 
bit larger than the one obtained for the drawings without coordinate grids ($2\fdg9$).
Note that proper motions in latitude are much smaller and extremely
rarely exceed $0\fdg1/$d. They do not noticeably contribute to $\sigma$.

\footnotetext[2]{http://solarscience.msfc.nasa.gov/SunspotCycle.shtml}

Figure~\ref{butterfly} shows the latitude--time distribution (butterfly 
diagram) of all sunspots measured in Schwabe's drawings. The patterns
formed by the four cycles observed do not show any peculiarities at 
first glance. {The separation of the two hemispheres is less
distinct than in the butterfly diagram of the RGO/USAF data set. This
is mostly due to the larger positional errors in the Schwabe
data, and to a lesser extent due to the fact that the RGO/USAF data are
average group positions while the Schwabe data contain individual spots
which introduce an additional intrinsic scatter to the plot.}

There were some periods in which the spot 
latitudes $b$ were very high, $|b|>50^\circ$. These were in August 1836, when spot latitudes exceeded 
$60^\circ$, in 1839, in the middle of Cycle~8, when latitudes exceeded $50^\circ$, and
in April 1854 when an individual spot was south of $-50^\circ$ at the
end of Cycle~9. When inspecting the apparent motion of the spots
across the disk, we noticed that the coordinate system given in
the drawings was not properly aligned. A total of 16~drawings have
therefore been analysed using the rotational matching of Section~\ref{bayesian}.
This method led to much lower latitudes for the first two periods
mentioned. The matching of the last period in 1854 (a single 
spot over seven days) did not deliver sharp probability density 
distributions and was discarded. April~24, 1854 with the exceptional 
latitude was removed from the data base. Most of these problematic 
drawings were actually not made by Schwabe, but by other persons. The butterfly
diagram also shows unusual latitudes in June 1846. Inspection of the
drawings shows, however, that the spot motion is consistent with the
alignment of the drawings. We have not altered these measurements in
the data base.

How likely are extreme latitudes?
The RGO/USAF data contain minimum and maximum group latitudes of 
$-59\fdg5$ and $59\fdg7$, respectively, according to the data base 
as of 2013 April~1\footnote[3]{http://solarscience.msfc.nasa.gov/greenwch.shtml}. 
Since these are average spot positions of a given group, the actual 
maximum and minimum latitudes of individual spots will be another 
few degrees towards the poles. A total of 14~sunspot groups have 
$|b|\geq 50^\circ$ in about 240,000~lines of data over almost 140~years 
in the RGO/USAF data base. The Schwabe measurements delivered 46~cases 
with $|b|\geq 50^\circ$ among about 135,000~lines of data, with extreme 
cases between $-52\fdg8$ and $56\fdg0$. There are relatively fewer high-latitude 
spots appearing in the RGO/USAF data than in Schwabe's data, but the extrema
are comparable.

\section{Summary}

We provide a set of about 135,000~sunspot positions and
sizes measured on drawings by Samuel Heinrich Schwabe in 
the period of November 5, 1825 to December 29, 1867. The data base can be obtained
from the web site of the corresponding author\footnote[4]{http://www.aip.de/Members/rarlt/sunspots}.
The accuracy of the sunspot positions appears to be between
three and four degrees in the heliographic coordinate
system near the disk center. We also include all verbal
reports on spotless days in the data base, so the file can
also be used for studies of the activity. The data also contain
an estimate of the individual spot sizes. They are given in
12~classes and should not be linearly scaled to physical areas.

The positions were obtained using (i) the coordinate system
drawn by Schwabe, if available, (ii) a rotational matching
with adjacent days if no coordinate system is given, and (iii)
an assumed alignment of the drawings with the horizontal
system, if (i) and (ii) were not applicable, which was the case predominantly in
the beginning of the observing period.

Note that we publish the first version of the data base here.
The data file may be updated at some time in the future if
errors emerge or the verbal information provides changes in
the interpretation of the drawings (most likely concerning
the clock times).

In the future, we intend to utilize also the information on spot 
evolution given in the verbal reports of Schwabe which are not 
accompanied by drawings. These improve the information
on the life-time of spots, since Schwabe carefully noted when 
spots disappeared and new spots appeared.

The potential of much less accurate drawings from the 18th century
has been demonstrated by Arlt \& Fr\"ohlich (2012) who determined
the differential rotation of the Sun based on the observations
by Johann Staudacher. The more careful drawings by Schwabe will
provide us with numerous quantitative results on four solar cycles
in the 19th century.

\section*{Acknowledgments}
RL thanks the V\"ais\"al\"a Foundation for financial support.
ND thanks Deutsche Forschungsgemeinschaft for the support
in grant Ar 355/7-1. We sincerely thank the Royal Astronomical
Society for their permission to digitize and utilize the original
manuscripts by Schwabe. The authors are very grateful to Anastasia 
Abdolvand, Luise Dathe, Jennifer Koch,
Sophie Antonia Penger, Clara Ricken and Christian Schmiel 
for their help with the measurements. We are highly indebted
to Hans-Erich Fr\"ohlich for providing his computer code for
Bayesian parameter estimations.

\bsp

\label{lastpage}

\end{document}